\documentclass[a4paper]{article}
\usepackage{amsmath}
\usepackage[english]{babel}
\usepackage{latexsym}
\usepackage{amssymb}
\usepackage{amscd}
\usepackage{amsgen,amstext,amsbsy,amsopn}

\newcommand{\version}{July 10, 2007}
\numberwithin{equation}{section}
\newcommand{\bdi}{\begin{displaymath}}
\newcommand{\edi}{\end{displaymath}}
\newcommand{\beqn}{\begin{eqnarray}}
\newcommand{\eeqn}{\end{eqnarray}}
\newcommand{\bay}{\begin{array}{c}}
\newcommand{\eay}{\end{array}}
\newcommand{\ben}{\begin{enumerate}}
\newcommand{\een}{\end{enumerate}}
\newcommand{\beq}{\begin{equation}}
\newcommand{\eeq}{\end{equation}}
\newcommand{\bmlt}{\begin{multline}}
\newcommand{\emlt}{\end{multline}}

\newcommand{\diff}{\mathrm{d}}

\newcommand{\err}{\mathbb{R}^2}

\newcommand{\gpfu}{\mathcal{E}_{V}^{\mathrm{GP}}}

\newcommand{\gpf}{\mathcal{E}^{\mathrm{GP}}}
\newcommand{\gpd}{\mathcal{D}^{\mathrm{GP}}}
\newcommand{\gpe}{E^{\mathrm{GP}}_{\varepsilon}}

\newcommand{\gpeu}{E^{\mathrm{GP}}_{\varepsilon, V}}

\newcommand{\gpm}{\Psi_{\varepsilon}^{\mathrm{GP}}}

\newcommand{\gpmu}{\Psi_{\varepsilon, V}^{\mathrm{GP}}}

\newcommand{\tgpm}{\tilde{\Psi}_{\vare}^{\mathrm{GP}}}
\newcommand{\tff}{\mathcal{E}^{\mathrm{TF}}}
\newcommand{\tfd}{\mathcal{D}^{\mathrm{TF}}}
\newcommand{\tfe}{E^{\mathrm{TF}}}
\newcommand{\tfm}{\rho^{\mathrm{TF}}}
\newcommand{\rtf}{R_{\mathrm{in}}}
\newcommand{\rmax}{R_{\mathrm{m}}}
\newcommand{\xin}{x_{\mathrm{in}}}
\newcommand{\xout}{x_{\mathrm{out}}}

\newcommand{\detfin}{\mathcal{T}_{\varepsilon}^{\mathrm{in}}}
\newcommand{\detfout}{\mathcal{T}_{\vare}^{\mathrm{out}}}
\newcommand{\detf}{\mathcal{T}_{\varepsilon}}

\newcommand{\bi}{\mathcal{B}_{\varepsilon}^j}

\newcommand{\trial}{\Psi_{\vare}}
\newcommand{\trialultra}{\tilde{\Psi}_{\xi, \vare}}
\newcommand{\cut}{\chi_{\varepsilon}}
\newcommand{\latt}{\mathcal{L}}
\newcommand{\spac}{\ell_{\varepsilon}}
\newcommand{\modu}{\rho_{\varepsilon}}
\newcommand{\phase}{g_{\varepsilon}}
\newcommand{\const}{C_{\omega_0}}
\newcommand{\constd}{C_{\omega_0,\delta}}
\newcommand{\constultra}{C_{\omega_1}}

\newcommand{\chem}{\mu_{\varepsilon}}

\newcommand{\chemtf}{\mu^{\mathrm{TF}}}
\newcommand{\tchemtf}{\tilde{\mu}^{\mathrm{TF}}}

\newcommand{\tfea}{E_{\varepsilon}^{\mathrm{TF}}}
\newcommand{\tfma}{\rho_{\varepsilon}^{\mathrm{TF}}}

\newcommand{\magnp}{\vec{A}_{\varepsilon}}
\newcommand{\vare}{\varepsilon}
\newcommand{\rf}{R_{\mathrm{out}}}
\newcommand{\rfc}{R_{\mathrm{out},c}}
\newcommand{\lam}{\Lambda}
\newcommand{\bol}{\mathcal{B}_{2\rf}}
\newcommand{\p}{\partial}
\newcommand{\sm}{\setminus}
\newcommand{\mc}{\mathcal}
\newcommand{\dens}{\tilde{\rho}^{\mathrm{TF}}}

\newcommand{\eneg}{\tilde{\mathcal{E}}^{\rm{GP}}}
\newcommand{\enef}{\tilde{\mathcal{E}}^{\rm{TF}}_{\varepsilon}}

\newtheorem{teo}{Theorem}[section]
\newtheorem{pro}{Proposition}[section]

\newtheorem{cor}{Corollary}[section]

\newenvironment{proof1}[1]{\mbox{} \newline \textbf{Proof of #1} \newline}{\begin{flushright} $ \Box $ \end{flushright}}
\newenvironment{proof}{\emph{Proof:}}{\begin{flushright} $ \Box $ \end{flushright}}

\begin{document}

\markboth{\scriptsize{CDY \version}}{\scriptsize{CDY \version}}

\title{Rapidly Rotating Bose-Einstein Condensates\\ in Homogeneous Traps}
\author{\hspace{-.2 cm} M. Correggi${}^{a}$
, T. Rindler-Daller${}^{c}$,\\ J. Yngvason${}^{b,c}$\\
\normalsize\it \hspace{-.5 cm}\\
\hspace{-.5 cm}\normalsize\it ${}^{a}$ Scuola Normale Superiore
SNS,    \\ \normalsize\it Piazza dei Cavalieri 7, 56126 Pisa, Italy \\ \normalsize\it
    ${}^{b}$ Erwin Schr{\"o}dinger
Institute for Mathematical Physics,\\ \normalsize\it
Boltzmanngasse 9, 1090 Vienna, Austria\\${}^{c}$\normalsize\it
Fakult\"at f\"ur Physik, Universit{\"a}t Wien,\\ \normalsize\it
Boltzmanngasse 5, 1090 Vienna, Austria}

\date{\version}

\maketitle

\begin{abstract}
We extend the results of a previous paper on the Gross-Pitaevskii description of rotating Bose-Einstein
condensates in two-dimensional traps to confining potentials of the
form $V(r) = r^s$, $2<s <\infty$. Writing the coupling constant
 as $1/\varepsilon^2$ we study the
limit $\varepsilon \to 0$. We derive rigorously the leading
asymptotics of the ground state energy and the density profile when
the rotation velocity $\Omega$ tends to infinity as a power of
$1/\varepsilon$. The case of asymptotically
homogeneous potentials is also discussed.

\vspace{0,5cm}

\end{abstract}

\section{Introduction}\label{Introduction}

In a previous  investigation of  rapidly rotating Bose-Einstein condensates in two-dimensional anharmonic traps  \cite{CDY} we considered the case of a  \lq flat' trap with a rigid boundary confining the condensate to a disk of  finite radius. The present paper is a sequel to this work, extending the results to homogeneous trap potentials of the form $V(r) = r^s$ with $2
< s < \infty$. The flat trap corresponds to the limiting case $s \to
\infty$. For $s<\infty$  the system is no longer
confined to a bounded region when the coupling constant  tends to infinity and a suitable scaling of the variables is necessary to
obtain a well-defined limit.  This gives rise to additional features and requires some modifications that are dealt with in the present paper. As in \cite{CDY}
we identify three different parameter regimes depending on the way the
rotation velocity is scaled with the interaction.

The present study is carried out strictly within the Gross-Pitaevskii (GP) framework in contrast to the recent paper \cite{BCPY} where  the main emphasis  is on many-body aspects and the GP description is an auxiliary tool. The trap potentials considered there are three dimensional and not necessarily rotationally symmetric. A two-dimensional potential of the form $r^s$  is an instructive special case that can be analyzed in more detail than the general case. 

As in \cite{CDY}, where a discussion of the general context and an
extensive list of references can be found, the starting point is the
two-dimensional Gross-Pitaevskii energy functional for the wave
function $\Psi$ of the condensate which in our case can be written
as \beq
    \label{gpf original}
    \hat{\mc{E}}^{\rm{GP}}[\Psi] \equiv \int_{\mathbb{R}^2} \diff \vec{r} \:
\left\{ |\nabla \Psi|^2 + r^s |\Psi|^2 - \Omega(\vare) \Psi^* L
\Psi + \frac{|\Psi|^4}{\vare^2}
      \right\}.
\eeq
Here $L$ is the third component of the angular momentum
(i.e. $L=-i\p /\p \vartheta$ in polar coordinates
$(r,\vartheta)$), $\Omega(\vare)$ the angular velocity and
$\vare$ is a non-negative, small parameter. The wave function is normalized so that
$\int_{{\mathbb R}^2} |\Psi|^2=1$. Units have been chosen so that $\hbar=2m=1$, where $m$ is the particle mass,  and such that the coefficient in front of $r^s$ is simply 1.

If  $s < \infty$ the condensate  spreads out indefinitely  in the Thomas-Fermi (TF) limit $\vare \to 0$ and the density $|\Psi|^2$ tends uniformly to zero. Non-trivial results can be obtained, however, by rescaling all lengths by an $\vare$-dependent factor. We write $\vec{r} \equiv k\vec{r}', \Psi(\vec{r}) \equiv \Psi'(\vec{r}')/k$ with $k \equiv \vare^{-2/(s+2)} $ and define
\beq
    \label{scaling}
    \gpf[\Psi'] \equiv \vare^{-\frac{4}{s+2}} \hat{\mc{E}}^{\rm{GP}}[\Psi].
\eeq
The functional $\gpf$ is the one we shall study. Dropping the primes of the arguments it is explicitly given by
\beq
    \label{gpf}
    \gpf[\Psi] = \int_{\mathbb{R}^2} \diff \vec{r} \: \left\{ |\nabla \Psi|^2 - \omega(\vare) \Psi^* L \Psi+ \frac{|\Psi|^2}{\vare^2}(r^s + |\Psi|^2)  \right\},
\eeq
where the scaled angular velocity is given by
\beq
    \label{skalang}
    \omega(\vare) \equiv \vare^{-\frac{4}{s+2}} \Omega(\vare).
\eeq
The functional \eqref{gpf} is defined on the domain
\bdi
    \gpd \equiv \{\Psi \in H^1(\mathbb{R}^2) \: | \: r^s |\Psi|^2 \in
    L^1(\mathbb{R}^2)\}.
\edi
We set
\beq
    \gpe \equiv \min_{\Psi \in \gpd, \| \Psi \|_2 = 1} \gpf[\Psi]
\eeq
and denote by $\gpm$ a corresponding minimizer, which may not be unique \cite{seir}.

In the following we study the leading order asymptotics of the ground state energy and density for the functional $\gpf[\Psi]$ as $ \vare \to 0 $. The ground state behavior for the original GP functional \eqref{gpf original} follows by scaling: If we set $ \hat{E}^{\mathrm{GP}}_{\varepsilon} \equiv \inf \hat{\mc{E}}^{\rm{GP}}[\Psi] $ and denote by $ \hat{\Psi}_{\varepsilon}^{\mathrm{GP}} $ any ground state of \eqref{gpf original}, one has
\beq
    \hat{E}^{\mathrm{GP}}_{\varepsilon} = \vare^{\frac{4}{s+2}} \gpe  \hspace{1,5cm} \hat{\Psi}_{\varepsilon}^{\mathrm{GP}}(\vec{r}) = \vare^{\frac{2}{s+2}} \gpm \left( \vare^{\frac{2}{s+2}} \vec{r} \right).
\eeq

Note that the `flat' trap case studied in \cite {FB} and \cite{CDY} can be formally obtained by taking the limit $s \to \infty$ of \eqref{gpf}: In this limit the scaling factor $\vare^{-2/(s+2)}$ converges to 1, the external potential to $ \infty $ for $ r > 1 $ and to 0 for $ r < 1 $,  and the rescaled angular velocity $ \omega(\varepsilon) $ to $ \Omega(\varepsilon) $. The formal limit $ s \to \infty $ corresponds to a `flat' trap with Dirichlet conditions at the boundary rather than the Neumann conditions considered in \cite{CDY}. As noted in \cite{CDY} both boundary conditions give the same results in the TF limit $\varepsilon\to 0$.

As in \cite{CDY} we rewrite the GP functional in the form
\beq
\label{gpfmag} \gpf[\Psi] = \int_{\mathbb{R}^2} \diff \vec{r} \: \left\{
\left| \left( \nabla - i \magnp \right) \Psi \right|^2 +
\frac{|\Psi|^2}{\vare^2}(r^s+|\Psi|^2) -
    \frac{\omega(\vare)^2 r^2 | \Psi |^2}{4} \right\}
\eeq with the vector potential \beq
    \label{magnp}
    \magnp(\vec{r}) \equiv \frac{\omega(\vare)}{2} \vec{e}_z \times
    \vec{r}
\eeq
 where $\vec{e}_z$ is the unit vector in $z$-direction. The behavior of  the GP functional  as $\vare \to 0$ depends on
 the way the  angular velocity $\omega$ scales as a function of  $\varepsilon$. We distinguish
 three cases:  $\omega\ll 1/\varepsilon$, $\omega\sim 1/\varepsilon$ and $\omega\gg 1/\varepsilon$. It
 is convenient to write $\omega=\omega_0/\varepsilon$; the three cases then correspond
 to $\omega_0\ll 1$, $\omega_0\sim 1$ and $\omega_0\gg 1$. In the next section we discuss
  for fixed $\omega_0$ the TF functional that is obtained from the GP functional by dropping the first
   (kinetic) term  in  \eqref{gpfmag}. This functional and its limits for $\omega_0\to 0$ and $\omega_0\to\infty$ describe the asymptotics of \eqref{gpfmag} for $\varepsilon\to 0$ as summarized in Section 3. In Section 4.1 we present the proofs for
the regimes $\omega\ll 1$ and $\omega \sim 1/\vare$ and in  Section 4.2  for $\omega \gg 1/\vare$.

Since  the length scale  $\varepsilon^{-2/(s+2)}$ tends to infinity in the TF limit, it is clear that for leading order calculations only the asymptotic behavior of the confining trap potential for large arguments matters. In the last section we indicate how our proofs can be extended to include potentials $V(r)$ that are asymptotically homogeneous in the sense of \cite{LSY}, i.e.,  such that for $\lambda\geq 1$
\beq
	\label{homogeneous}
	\left| \lambda^{-s} V(\lambda r) - r^s \right| \leq c \lambda^{-\kappa} \left( 1+ r^s \right)
\eeq
for some constants $\kappa,c>0$, uniformly in $r\in\mathbb R^+$.

\section{The TF functional and its properties}
The TF functional depends on the density alone and is for fixed $\omega_0$ defined as
\beq
    \label{tff2}
        \tff[\rho] \equiv \int_{\mathbb{R}^2} \diff \vec{r} \: \left\{ \rho (r^s + \rho) - \frac{\omega_0^2 r^2 \rho}{4} \right\}
\eeq
on the domain
\beq
        \tfd \equiv \left\{ \rho \in L^2(\mathbb{R}^2) \: | \: \rho \geq 0, r^s \rho \in L^1(\mathbb{R}^2) \right\}.
\eeq
By  standard methods there is  a unique minimizer
\beq
    \label{tfmhom}
    \tfm(r) \equiv \frac{1}{2}\left[\chemtf - r^s + \frac{\omega_0^2 r^2}{4}\right]_+
\eeq
where the chemical potential $\chemtf$ is fixed by $ \| \tfm \|_1 = 1$. The ground state energy associated with \eqref{tfmhom} is
\beq
    \tfe \equiv \inf_{\rho \in \mc{D}^{TF}, \| \rho \|_1 = 1} \tff[\rho]=\tff[\rho^{\rm TF}].
\eeq
Since $ - r^s + \omega_0^2r^2/4 \to - \infty $ as $ r \to \infty $, due to the condition $ s>2 $, the minimizer \eqref{tfmhom}  is always compactly supported, i.e., $ \rm{supp}(\tfm) \subset \mathcal{B}_R $, for some $ R < \infty $ depending on $\omega_0$ where $ \mathcal{B}_{R} $ denotes a two-dimensional ball centered at the origin with radius  $ R $.
\newline
As in the case of a flat trap the density $\rho^{\rm TF}$ develops a `hole' (a disk centered at the origin where the density \eqref{tfmhom} vanishes) when $\omega_0$ exceeds a certain critical value $\omega_{0,c}$. Both the outer radius $\rf$ of the support of $\tfm$  and the inner radius $\rtf$ (if present)  increase with $\omega_0$ as discussed below. In the flat trap only $\rtf$  increases with
$\omega_0$ while $\rf$ is fixed from the outset.
\\
The chemical potential $\chemtf$ depends also on $\omega_0$ and because $-r^s +
\omega_0^2 r^2/4$ is monotone increasing in $\omega_0$, it is
clear that $\chemtf$ is monotone decreasing in $\omega_0$ (see
also \eqref{form} and \eqref{form2}). For small  $\omega_0$ we have
$\chemtf > 0$, whereas  $ \chemtf $ vanishes and changes sign at $\omega_0 = \omega_{0,c}$.

\subsection{Support of $ \tfm $ for $ \omega_0 = \mathrm{const.} $}
\label{support finite}

In order to study the support of $ \tfm $ it is convenient to consider the function
$   f(z) \equiv \chemtf - z^{s/2} + {(\omega_0^2/4)z}$
where $ z \equiv r^2\geq 0 $, so that $ \tfm = f(r^2)_+ $. Since
${\diff^2 f}/{\diff z^2} = - {s(s-2)z^{\frac{s-4}{2}}}/{4} < 0$
for $ z \in (0,\infty) $, the function $ f $ is strictly concave for $ z > 0$. Moreover $ f(0) = \chemtf  $ and $ \lim_{z \to \infty} f(z) = - \infty $.
Hence, if $\chemtf > 0$, i.e., $\omega_0 < \omega_{0,c}$, then $ f(0) > 0 $ and there exists a unique  $ z_{\mathrm{out}} > 0 $ such that $ f(z_{\mathrm{out}}) = 0 $. The support of $ \tfm $ in the radial coordinate is the interval $[0, \rf] $, with $ \rf \equiv \sqrt{z_{\mathrm{out}}} $.
In the opposite case $\chemtf < 0$, i.e., $\omega_0 > \omega_{0,c}$, we have  $ f(0) < 0 $ but $ \sup f > 0 $ (since $\int \tfm >0$) and concavity implies the existence of two positive solutions of $ f(z) = 0 $. The support of $ \tfm $ in the radial coordinate is then an  interval $[\rtf,\rf] $ for some $ 0 < \rtf < \rf $.  Note also that
\beq
\label{fprime}
 0<f'(\rtf^2)= -\frac s2\rtf^{s-2}+\frac{\omega_0^2}4\quad\hbox{and}\quad 0>f'(\rf^2)= -\frac s2\rf^{s-2}+\frac{\omega_0^2}4.
\eeq

From the $L^1-$normalization of $ \tfm $ and
\beq
\label{mu}
    \chemtf = \rf^s - \frac{\omega_0^2\rf^2}{4}
\eeq
we get for $\omega_0 < \omega_{0,c}$,
\beq
    \label{radtf}
        \frac{s\rf^{s+2}}{2(s+2)} - \frac{\omega_0^2\rf^4}{16} = \frac{1}{\pi},
\eeq
whereas for $\omega_0 > \omega_{0,c}$ we have
\beq
    \label{radtf2}
    \chemtf = \rtf^s - \frac{\omega_0^2\rtf^2}{4} = \rf^s - \frac{\omega_0^2\rf^2}{4},
\eeq
\beq
    \label{radhole}
    \frac{s(\rf^{s+2} - \rtf^{s+2})}{2(s+2)} - \frac{\omega_0^2(\rf^4-\rtf^4)}{16} = \frac{1}{\pi}.
\eeq
The radii $\rf$ and $\rtf$ are determined by solving \eqref{radtf}, or \eqref{radtf2} together with \eqref{radhole}. While explicit formulas can in general not be given, the
outer radius $ \rfc $ for $ \chemtf = 0 $ and the critical angular velocity $ \omega_{0,c} $ for the creation of a hole are easily obtained from \eqref{mu} and \eqref{radtf}:
\beq
    \rfc = \left(\frac{\omega_{0,c}}{2}\right)^{\frac{2}{s-2}}
\eeq
with
\beq
    \omega_{0,c} = 2\left[\frac{4(s+2)}{\pi(s-2)}\right]^{\frac{s-2}{2(s+2)}}.
\eeq
In the `flat' trap case, i.e., for $ s \to \infty $, the expression for $ \omega_{0,c} $ simplifies to $ 4/\sqrt{\pi}$ as in \cite{CDY}.

We  now show  that $ \chemtf $ is monotonically decreasing and $ \rtf $ and $ \rf $ monotonically increasing as $ \omega_0 $ increases.
Consider first the case $ \omega_0 < \omega_{0,c} $, i.e., $\mu^{\rm TF}>0$. Differentiating the normalization equation $\int\rho^{\rm TF}=1$  with respect to
$ t \equiv \omega_0^2/4$, using \eqref{mu}, gives
\beq\label{form}
\partial\mu^{\rm TF}/\partial t=-\rf^2/2<0.
\eeq
Differentiating \eqref{radtf} gives
\beq\label{partialrf}
{\partial \rf}/{\partial t}=(s\rf^{s-2}-2t)^{-1}\rf/ 2
\eeq
and hence $\partial \rf/\partial t>0$ because of \eqref{fprime}.

In the case $ \omega_0 > \omega_{0,c} $, we again differentiate the normalization condition for $\rho^{\rm TF}$, this time using \eqref{radtf2}, and obtain
\beq\label{form2}
\partial\mu^{\rm TF}/\partial t=-(\rf^2+\rtf^2)/2<0.
\eeq
 Moreover, by taking the derivative of \eqref{radtf2} w.r.t. $ t $, we have
\bdi
    \frac{\partial \chemtf}{\partial t} = (s\rtf^{s-1}-2t\rtf) \frac{\partial \rtf}{\partial t} - \rtf^2 = (s\rf^{s-1}-2t\rf) \frac{\partial \rf}{\partial t} - \rf^2,
\edi
which, combined with \eqref{form2}, yields the inequalities
\bdi
    (s\rtf^{s-1}-2t\rtf) \frac{\partial \rtf}{\partial t} < 0 \quad\hbox{and}\quad (s\rf^{s-1}-2t\rf) \frac{\partial \rf}{\partial t} > 0.
\edi
By \eqref{fprime} this implies
\beq
    \frac{\partial \rtf}{\partial t} > 0 \quad\hbox{and}\quad\frac{\partial \rf}{\partial t} > 0.
\eeq

Altogether, we have thus seen that $\tfm$ has
compact support contained in $\mathcal{B}_{\rf} $. If $ \omega_0 < \omega_{0,c} $, the support coincides with $ \mathcal{B}_{\rf} $, while, for  $ \omega_0 > \omega_{0,c} $, it  is the annulus $\{\vec r:\, \rtf\leq r\leq\rf\}$. Both $\rtf$ and $\rf$ grow as $\omega_0$ increases.

\subsection{The quartic trap}

Let us consider for illustration the quartic trap $V(r)=r^4$. The
critical angular velocity and radius are given by
\bdi
    \omega_{0,c} = 2\left(\frac{12}{\pi} \right)^{1/6}, \hspace{1,5cm} \rfc = \left(\frac{12}{\pi} \right)^{1/6}.
\edi The equations for the inner and outer radii $\rtf$ and $\rf$
reduce to equations of third order for the squares of the radii.  If
$\omega_0 \leq \omega_{0,c} $ the chemical potential and the outer
radius are \bdi
        \chemtf = \frac{1}{64}\left\{\frac{1}{2}\left(\frac{q(\omega_0)}{\pi}\right)^{1/3} +
        \frac{\omega_0^4}{2} \left(\frac{\pi}{q(\omega_0)}\right)^{1/3} - \frac{\omega_0^2}{2}\right\}^2 -
        \frac{\omega_0^4}{64}
\edi
and
 \bdi
  \rf = \frac{1}{4}\left\{\left(\frac{q(\omega_0)}{\pi}\right)^{1/3}
  + \omega_0^4 \left(\frac{\pi}{q(\omega_0)}\right)^{1/3} +
  \omega_0^2\right\}^{1/2}
   \edi
 with $q(\omega_0) \equiv 6144 + \pi \omega_0^6 + 64
   \sqrt{3}\sqrt{3072+\pi \omega_0^6}$.
   If $\omega_0 > \omega_{0,c} $, we obtain
 \bdi
        \chemtf =
        \frac{1}{4}\left(\frac{12}{\pi}\right)^{2/3}-\frac{\omega_0^4}{64},
\edi
 \bdi
  \rtf =
  \sqrt{\frac{\omega_0^2}{8}-\frac{1}{2}\left(\frac{12}{\pi}\right)^{1/3}},~~
  \rf =
  \sqrt{\frac{\omega_0^2}{8}+\frac{1}{2}\left(\frac{12}{\pi}\right)^{1/3}}.
  \edi
(Note that the two expressions for $ \chemtf $ and $\rf$ are equal
for $ \omega_0 = \omega_{0,c} $ whereas $\rtf = 0$.) For $\omega_0 >
\omega_{0,c} $ the TF minimizer is \bdi
        \tfm(r) = \left[\frac{\omega_0^2}{8} \left(r^2-\frac{\omega_0^2}{16}\right) - \frac{r^4}{2} + \frac{1}{8}\left(\frac{12}{\pi}\right)^{2/3} \right]_+ .
\edi

\subsection{Support of $ \tfm $ for $ \omega_0 \to \infty $}
\label{support infinite}

The radius $ \rmax $ at which
the density is maximal can be explicitly calculated from
\eqref{tfmhom}:
\beq
    \label{rmax}
    \rmax \equiv \left(\frac{\omega_0^2}{2s}\right)^{\frac{1}{s-2}}
\eeq
It is clear that $ \rtf < \rmax<\rf$ and all radii tend to infinity if $\omega_0\to\infty$.
We shall now show that $\rf-\rtf$  tends to zero in this limit. For $s>4$ also $\rf^2-\rtf^2$
tends to zero and the density  therefore to infinity.

It is convenient to scale all lengths by using $ \rmax $ as a unit, i.e, to write  $r = \rmax x$. The scaled TF minimizer $ \dens(x) \equiv \rmax^2 \tfm(r) $ is
\beq
    \label{skaldens}
    \dens(x) \equiv \frac{1}{2}\left(\frac{\omega_0^2}{2s}\right)^{\frac{s+2}{s-2}} \left[\tchemtf - x^s + \frac{s x^2}{2}\right]_+,
\eeq
with the scaled chemical potential
\beq
        \tchemtf \equiv \left(\frac{\omega_0^2}{2s}\right)^{-\frac{s}{s-2}} \chemtf.
\eeq
We also denote $ \xin \equiv \rtf/ \rmax $ and $ \xout \equiv \rf / \rmax $, so that $ 0 \leq \xin < 1 < \xout $ and the maximum of $ \dens $ is attained at $ x = 1 $.
\newline
In the same way as  \eqref{form2} was derived we have
\beq
    \label{forma}
    \frac{\p \tchemtf}{\p \omega_0^2} = - \left(\frac{\omega_0^2}{2s}\right)^{-\frac{2s}{s-2}} \frac{(s+2) }{\pi s(s-2)(\xout^2-\xin^2)} < 0
\eeq
so that $ \tchemtf $ is a  decreasing function of $ \omega_0 $. Moreover since $ \dens(1) > 0 $, one has the bound
\beq
    -\tchemtf < \frac{s}{2}-1.
\eeq
Defining
\beq
h(x)\equiv  x^s - (s/2) x^2+(s/2)-1,
\eeq
the scaled radii $\xin$ and $\xout$ are determined by
\beq
h(\xin)=h(\xout)
\eeq
together with the normalization condition for $\dens$ which expressed in terms of $ h$ is
\beq\label{hnorm}
\frac{h(\xin)}2\left({\xout}^2-{\xin^2}\right)-\int_{\xin}^{\xout} h(x)x dx=\pi^{-1}\left(\frac{\omega_0^2}{2s}\right)^{-\frac{s+2}{s-2}}.
\eeq
The right hand side of \eqref{hnorm} tends to zero as $\omega_0\to\infty$ and the integral on the left hand side is always strictly less than the first term for $\xin$ strictly less than $\xout$. Since $\xin<1<\xout$ and $h$ is continuous with $h(x)=0$ only for $x=1$ it is clear that $\xin$ and $\xout$ must both tend to 1 as $\omega_0\to\infty$ and by the normalization the density $\dens$ approaches a delta function concentrated on the circle with radius 1. Note also that
\beq
	\label{mutildelim}
	h(\xin)=h(\xout)=\tilde\mu^{\rm TF}+\frac s2-1
\eeq
so that
\beq 
	\tilde\mu^{\rm TF} \underset{\omega_0\to\infty}{\longrightarrow} 1- \frac{s}{2}.
\eeq

In order to estimate the rate of the convergence of the density to a delta function we make a  Taylor
expansion of  $h(x)$  around $x=1$: \beq\label{taylor}
h(x)=\frac{1}{2}s(s-2)(1-x)^2+O(|1-x|^3). \eeq Writing $x_{\rm
in}=1-\delta+o(\delta)$, $x_{\rm out}=1+\delta+o(\delta)$, where
$\delta$ is the deviation from 1 to leading order in the small
parameter on the right hand side of \eqref{hnorm}, the normalization
condition \eqref{hnorm}  gives \beq
\delta=\left(\frac3{s(s-2)}\right)^{1/3}\left(\frac{\omega_0^2}{2s}\right)^{-\frac{s+2}{3(s-2)}}.
\eeq Multiplying $\delta$ with $R_{\rm m}\sim \omega_0^{2/(s-2)}$ we
see that \beq (R_{\rm out}-R_{\rm in})\sim\omega_0^{-2(s-1)/3(s-2)}.
\eeq Thus also the original  density $\rho^{\rm TF}$ is supported on
an annulus whose thickness tend to zero.  The area of the support is
\beq\pi(R_{\rm out}^2-R_{\rm in}^2)\sim \omega_0^{2(4-s)/3(s-2)}
\eeq which increases for $2<s<4$ but tends to zero for $s>4$.

\subsection{TF energy asymptotics for $\omega_0\to\infty$}
The scaled density $\tilde\rho^{\rm TF}$ is the minimizer of the scaled TF functional
\beq
    \label{TFfastrot}
    \tilde{\mathcal E}^{\rm TF}[\tilde{\rho}] \equiv \int_{\mathbb{R}^2} \diff\vec{x} \left[ \left( x^s - \frac{s x^2}{2} \right) \tilde{\rho} + \left( \frac{\omega_0^2}{2s} \right)^{-\frac{s+2}{s-2}} \tilde{\rho}^2 \right]
\eeq
with corresponding energy $\tilde E^{\rm TF}=\tilde{\mathcal E}^{\rm TF}[\tilde{\rho}^{\rm TF}]$.  As shown in the previous subsection  $\tilde\rho^{\rm TF}$ converges to a delta function on the unit circle as $\omega_0\to\infty$.
The behavior of the energy in this limit is given in the following proposition.
\begin{pro}[TF energy  for $\omega_0\to \infty$]
\label{TFenergyasympt}
\mbox{} \\
For $\omega_0 \to \infty$, 
\beq
    \label{tfentildeasympt}
    \tilde E^{\rm TF} = 1 - \frac{s}{2} + \mathcal{O}\left(\omega_0^{-\frac{4 (s+2)}{3(s-2)}}\right).
\eeq
\end{pro}
\begin{proof}
    The lower bound is simply obtained by neglecting the positive last term in \eqref{TFfastrot} and
     using the inequality $ x^s - sx^2/2 \geq 1 - s/2 $. For the upper bound we use a  trial function of the form
    \beq
    \label{trialdensity}
         \tilde{\rho}_{\xi}(x) \equiv \xi^{-1} j(\xi^{-1}(1-x^2))
    \eeq
    where $ j $ is a smooth non-negative function supported in $ [-1/2,1/2] $ satisfying the normalization $ \pi \int \diff r j(r) = 1 $ and   $ 0 < \xi < 1 $. One can easily estimate $ \| \tilde{\rho}_{\xi} \|_2^2 \leq C \xi^{-1} $, and exploiting  the Taylor expansion of $ x^s - sx^2/2 $ around $ x = 1 $ we have
    \bdi
        \int_{\mathbb{R}^2} \diff \vec{x} \: \left( x^s - \frac{s x^2}{2} \right) \tilde{\rho}_{\xi} \leq 1 - \frac{s}{2} + C' \xi^2
    \edi
    so that
    \beq
        \label{estimateTFultra}
        \tilde{\mc{E}}^{\rm TF}[\tilde{\rho}_{\xi}] \leq 1 - \frac{s}{2} + C' \xi^2 + C^{\prime\prime} \: \omega_0^{-\frac{2(s+2)}{s-2}} \xi^{-1}.
    \eeq
    Optimization with respect to the parameter $ \xi $ yields the desired upper bound.
    \end{proof}
\section{Main results}
\label{main results}

\subsection{The regime $\omega \ll 1/\vare$}

For $\omega \ll 1/\vare$, the GP ground state energy and density are approximated to the leading order by the corresponding quantities in the non-rotating case, exactly as in Proposition 2.3 in \cite{CDY}. The TF functional without rotation, i.e., for $ \omega_0 = 0 $, is given by
\bdi
    \tff_*[\rho] \equiv \int_{\err} \diff \vec{r} \: \rho(r^s+\rho).
\edi
We denote by $ \tfe_* $ its ground state energy and by
\bdi
    \tfm_*(r) \equiv \frac{1}{2}\left[\chemtf - r^s \right]_+
\edi
the corresponding minimizer.

\begin{pro}[GP energy and density asymptotics for $\omega\ll 1/\varepsilon$]
    \label{norotationhom}
    \mbox{} \\
    For any \( \omega(\varepsilon) \) such that \( \lim_{\varepsilon \to 0} \varepsilon \omega(\varepsilon) = 0 \)
    and for \( \varepsilon \) tending to zero,
        \beq
            \label{energynorothom}
            \varepsilon^2 \gpe = \tfe_* + o(1),
        \eeq
        \beq
            \label{convnorothom}
            \left\| |\gpm|^2 - \tfm_* \right\|_{L^2(\mathbb{R}^2)} = o(1).
        \eeq
\end{pro}

\begin{proof}
The lower bound for the ground state energy is actually trivial, since it is sufficient to neglect the first positive term in \eqref{gpfmag} to obtain
\bdi
    \varepsilon^2 \gpe \geq \tfe_* - C \varepsilon^2 \omega(\varepsilon)^2.
\edi
In order to get an appropriate upper bound we test the GP functional on the (real) GP minimizer for $ \omega = 0 $ and obtain $ \gpe \leq \gpe|_{\omega=0} $. The result is then a consequence of Lemma 2.3 in \cite{LSY}
and we get the bound (see Eq. (2.18) in \cite{LSY})
\beq
    \varepsilon^2 \gpe \leq \tfe_* + C \varepsilon^{2/3}.
\eeq
The density convergence is a simple corollary (see the proof of Proposition 2.3 in \cite{CDY} and Theorem 2.1 in \cite{LSY}).
\end{proof}

\subsection{The regime $\omega \sim 1/\vare$}

We now assume that $\omega(\vare) = \omega_0/\vare$ with $\omega_0
> 0$ a finite constant. The analogs of Theorem 2.1 and Corollary 2.1 in \cite{CDY} are the following:

\begin{teo}[GP energy asymptotics for $\omega\sim 1/\varepsilon$]
    \label{energyhom}
    \mbox{} \\
    For any \( \omega_0 > 0 \) and for \( \varepsilon \) tending to zero,
    \beq
            \varepsilon^2 \: \gpe = \tfe + \mathcal{O}(\varepsilon | \log\varepsilon |).
    \eeq
\end{teo}

\begin{cor}[GP density asymptotics for $\omega\sim 1/\varepsilon$]
    \label{profilehom}
    \mbox{} \\
    For any \( \omega_0 > 0 \) and for \( \varepsilon \) tending to zero,
    \beq
            \label{L1profilehom}
            \left\| | \gpm |^2 - \tfm \right\|_{L^2(\mathbb{R}^2)} = \mathcal{O}(\sqrt{\varepsilon |\log \varepsilon|}).
    \eeq
\end{cor}
The asymptotics of the energy and density for the original functional \eqref{gpf original}  quantities is then given by
\bdi
    \vare^{-\frac{2s}{s+2}} \hat{E}^{\mathrm{GP}}_{\varepsilon} = \tfe + \mathcal{O}(\varepsilon | \log\varepsilon |)  \quad\hbox{and}\quad  \vare^{-\frac{4}{s+2}} \left|\hat{\Psi}_{\varepsilon}^{\mathrm{GP}} \left( \vare^{-\frac{2}{s+2}} \vec{r} \right) \right|^2 \underset{\vare \to 0}{\longrightarrow} \tfm(r),
\edi
where the convergence of the density is in the norm topology of $ L^1(\mathbb{R}^2) $.

For $s < \infty$, the condensate is not confined to a bounded region
and $|\gpm|^2$ is a function supported on the whole of
$\mathbb{R}^2$. From Corollary \ref{profilehom}  it follows immediately that  $|\gpm|^2$ is small
outside the support of $ \tfm $ in $L^2$ norm, i.e., \beq
\label{outside support}
    \int_{\mathbb{R}^2\setminus \mathrm{supp}(\tfm)} \diff \vec{r} \: |\gpm|^4 = \mathcal{O}(\varepsilon |\log \varepsilon|)
\eeq but much more can be shown, namely that $ |\gpm| $ is pointwise
exponentially small outside the support of $ \tfm $:
\begin{teo}[Exponential smallness of the GP density,  $\omega\sim 1/\varepsilon$]
    \label{pom}
    \mbox{} \\
    For any $ \omega_0 > 0 $, $\vec r \in \detfout \equiv \{ \vec{r} \in \mathbb{R}^2 \: | \: r \geq \rf + \vare^{1/3} \} $ and for \( \varepsilon \) sufficiently small,
    \beq
            \label{denexp}
            |\gpm (\vec{r})|^2  \leq  \const \vare^{1/6} \sqrt{|\log \vare|} \exp \left[- \frac{\const^{\prime} \mathrm{dist}(\vec{r},\p \detfout)^2}{\vare^{5/6}} \right].
    \eeq
    Furthermore, for any $ \omega_0 > \omega_{0,c} $ the same
estimate holds for
     $ \vec{r} \in \detfin \equiv \left\{ \vec{r} \in \mathbb{R}^2 \: | \: r \leq \rtf - \varepsilon^{1/3} \right\} $ and $ \partial\detfout $ replaced with $ \partial\detfin $.
     \end{teo}

\subsection{The regime $\omega \gg 1/\vare$}

For convenience, in particular for the statement of Theorem \ref{expahom} below,  and comparison with \cite{CDY} we assume that $\omega$ increases as a power of $1/\varepsilon$, i.e., that $\omega(\vare) = \omega_1/\vare^{1+\alpha}$ with some
constants $\omega_1, \alpha > 0$. This means that we take
$\omega_0=\varepsilon\omega(\varepsilon)=\omega_1/\varepsilon^\alpha$.
 Theorem  \ref{energyahom} holds true for general $\omega_0(\varepsilon)=\varepsilon\omega(\varepsilon)\to\infty$  if   $\omega_1/\varepsilon^\alpha$ is replaced by $\omega_0(\varepsilon)$.

In the regime $\omega\gg 1/\varepsilon$ the limiting functional is
still given by \eqref{tff2}, but since  $ \omega_0 $ now depends on
$\varepsilon$ this is also the case for the TF ground state energy
and density. We thus use the notations $ \tfea $ and $ \tfma $.
Proposition \ref{TFenergyasympt} yields the ground state energy
asymptotics for the  functional $ \tfea $, i.e., \beq \label{etf}
    \vare^{\frac{2\alpha s}{s-2}} \tfea = \left( \frac{\omega_1^2}{2s} \right)^{\frac{s}{s-2}} \left( 1 - \frac{s}{2} \right) + \mathcal{O}\left(\vare^{\frac{4 \alpha(s+2)}{3(s-2)}}\right).
\eeq
The following Theorem describes the GP ground state energy asymptotics.

\begin{teo}[GP energy asymptotics for $\omega\gg 1/\varepsilon$]
    \label{energyahom}
    \mbox{} \\
        For any \( \omega_1, \alpha > 0 \) and \( \varepsilon \) tending to zero,
        \beq
        \label{energyaesthom}
            \varepsilon^{2} \: \vare^{\frac{2\alpha s}{s-2}} \: \gpe = \left( \frac{\omega_1^2}{2s} \right)^{\frac{s}{s-2}} \left( 1 - \frac{s}{2} \right)  + \mathcal{O} \left( \varepsilon^{\frac{4\alpha(s+2)}{3(s-2)}} \right) + \mathcal{O} \left( \vare \: \varepsilon^{\frac{\alpha(s+2)}{s-2}} \right).
        \eeq
\end{teo}

Note the occurrence of two remainders in \eqref{energyaesthom}: The first one,
of order $ \varepsilon^{\frac{4\alpha(s+2)}{3(s-2)}} $, is actually the expected optimal one,
 since it coincides with the (optimal) error term in \eqref{etf}. Therefore,
 as long as $ \alpha \leq \frac{3(s-2)}{s+2} $, the second term is just a higher order
 correction and the result is optimal as far as the order of the error term is concerned. However, for larger $ \alpha $ the leading correction in \eqref{energyaesthom} is given by the second error term  and it is due to the particular form of the trial function involved in the proof (see Section \ref{proofsultra}).

In order to state a pointwise estimate analogous to \eqref{denexp}, it is convenient to rescale the GP minimizer in the same way as when $\tilde\rho^{\rm TF}$ was obtained from $\rho^{\rm TF}$ by scaling. Thus we define (see also \eqref{rescaledultra})
\beq
    \tgpm(\vec{x}) \equiv \rmax \gpm(\rmax \vec{x})
\eeq
with $ \vec{x} \equiv \rmax^{-1} \vec{r} $. The scaled  minimizer $\tgpm$  is concentrated in a neighborhood of $ x = 1 $ and exponentially small  everywhere else:

\begin{teo}[Exponential smallness of the GP density, $\omega\gg 1/\varepsilon$]
    \label{expahom}
    \mbox{} \\
    Set
    \beq
        \beta \equiv \min \left[ \frac{4\alpha(s+2)}{3(s-2)} \:, \: 1 + \frac{\alpha(s+2)}{s-2} \right].
    \eeq
    For any $ \alpha, \omega_1 > 0 $, $\vec{r} \in \detf \equiv \{ \vec{r} \in \mathbb{R}^2 \: | \: |1-r| \geq \vare^{\beta/3} \} $ and for \( \varepsilon \) tending to zero,
        \beq
    \label{pointestultra}
            \left| \tgpm(\vec{x}) \right|^2 \leq  \constultra \vare^{\beta/6} \varepsilon^{-\frac{\alpha(s+2)}{s-2}} \exp \left[ - \frac{ \constultra^{\prime} \mathrm{dist}(\vec{x}, \partial \detf)^2}{\varepsilon^{\gamma}} \right]
        \eeq
    where
    \beq
        \gamma \equiv 1 - \frac{\beta}{3} + \frac{\alpha(s+2)}{s-2}.
    \eeq
    (Note that for both possible values of $ \beta $ the exponent $ \gamma $ is positive.) Furthermore the density $ | \tgpm(\vec{x}) |^2 $ converges in the sense of distributions to a Dirac delta function concentrated at $ x =1 $.
\end{teo}

\section{Proofs}

In this Section we prove the main results mentioned in Section \ref{main results}.

\subsection{The regime $\omega(\vare) \sim 1/\vare$}

We start by proving the ground state energy asymptotics and the other results will follow as simple corollaries. The proof is quite similar to the proof of Theorem 2.1 in \cite{CDY}: Like there, the main ingredient in the derivation of the upper bound for the energy  is a trial function with a large number of vortices while the differences are essentially contained in  a scaling argument.

\begin{proof1}{Theorem \ref{energyhom}}
	The lower bound is obtained again by simply neglecting the positive \lq magnetic' kinetic energy in \eqref{gpfmag}, namely \beq
        \label{lboundhom}
            \gpf[\Psi] \geq \frac{\tff[|\Psi|^2]}{\varepsilon^2} \geq \frac{\tfe}{\varepsilon^2}.
        \eeq
	To get an  upper bound we evaluate the GP functional on a trial function of the form
\beq
        \label{trialhom}
        \trial(\vec{r}) = c_{\varepsilon} \sqrt{\modu(r)} \: \cut(\vec{r}) \phase(\vec{r}),
\eeq
where $ \phase $ is a phase factor, $ \cut $ a  function that vanishes at the singularities of $ \phase $ and $ \modu $ a suitable regularization of $ \tfm $. More precisely $ \modu $ is defined as in Lemma 2.3 in \cite{LSY}, i.e., $ \modu \equiv j_\varepsilon\star \tfm $, with
        \beq
        \label{cut modulus}
            j_\varepsilon(r) \equiv \frac{1}{2\pi \varepsilon^2} \exp \left\{ -\frac{r}{\varepsilon} \right\}.
        \eeq
        Since $ \| j_{\varepsilon} \|_1 = 1 $, $ \sqrt{\modu} $ is $L^2-$normalized. It is also clear that $ \modu $ converges uniformly to $ \tfm $ as $ \varepsilon \to 0 $ and it is uniformly bounded in $ \varepsilon $, i.e., there exists a constant $ \const $ such that $ \modu \leq \const $. Furthermore, although $ \modu $ is not compactly supported, it is exponentially small in $ \varepsilon $ for $ \vec{r} $ sufficiently far from the support of $ \tfm $: For any $ \vec{r} \in \mathbb{R}^2 $, $ r > \rf $,
    \beq
    \label{exponential smallness}
            \modu(\vec{r}) = \frac{1}{2 \pi \varepsilon^2} \int_{r \leq \rf} \diff \vec{r}^{\prime} \: \exp \left\{ -\frac{\left| \vec{r} - \vec{r}^{\prime} \right|}{\varepsilon} \right\} \tfm(r^{\prime}) \leq \frac{1}{2\pi \varepsilon^2} \exp \left\{ -\frac{r-\rf}{\varepsilon} \right\}.
    \eeq
Moreover,  two different estimates for the gradient of $ \modu $ hold true: By using the fact that $ |\nabla j_{\varepsilon}| = \varepsilon^{-1} j_{\varepsilon} $, one can easily prove that $ \left| \nabla \modu \right|  \leq \varepsilon^{-1} \left| \modu \right| $, whereas by exploiting the regularity of $ \tfm $, i.e., $ \left\| \nabla \tfm \right\|_1 \leq \const $, one has $  \left\| \nabla \modu \right\|_1 \leq \const $. Using both estimates we immediately get the bound 
\beq\label{TFkin}\| \nabla \sqrt{\modu} \|_2^2 \leq \const/\vare .\eeq

The phase factor $g_{\vare}$ and the cutoff function $\cut$ are defined as in \cite{CDY}
by placing  vortices of degree 1  at the points of the square lattice
\beq
        \label{latticehom}
            \latt = \left\{ \vec{r}_j = (m \spac, n \spac), \: \: m,n \in \mathbb{Z} \: \Big| \: r \leq 2\rf - 2 \sqrt{2} \spac \right\}
        \eeq
with spacing $\spac = \delta \sqrt{\vare} $ for some $ \delta > 0$:
Using complex notation $ \zeta=x+iy$ for $\vec r=(x,y)\in\mathbb R^2$ we define
\beq
    \label{phase}
    g_{\vare}(\zeta) = \prod_{\zeta_j \in \latt} \frac{\zeta-\zeta_j}{|\zeta-\zeta_j|}
\eeq
\beq
            \cut(\vec{r}) =
            \left\{
            \begin{array}{ll}
                    1   &   \mbox{if} \:\:\:\: |\vec{r} - \vec{r}_j| \geq \varepsilon^{\eta}  \\
                    \mbox{} &   \mbox{} \\
                    \displaystyle{\frac{|\vec{r} - \vec{r}_j|}{\varepsilon^{\eta}}}   &   \mbox{if} \:\:\:\: |\vec{r} - \vec{r}_j| \leq \varepsilon^{\eta}
                \end{array}
            \right.
\eeq
for some \( \eta > 5/2 \).
Note that the vortex lattice $\latt$ has the same spacing as in \cite{CDY} but it is extended to cover the whole of the support of $ \tfm $. The number, $ N_{\vare} $,  of vortices  and the normalization constant $ c_{\vare} $ satisfy the bounds $ N_{\vare} \leq \constd/\vare $, due to \eqref{latticehom}, and $ 1 \leq c^2_{\varepsilon} \leq  1 + o(\varepsilon^4) $, since $ \cut \leq 1 $ and $ \eta > 5/2 $. By setting
\bdi
    \lam \equiv \bol \sm \bigcup_{j \in \latt} \bi,
\edi
where $\bi$ is a ball of radius $\varepsilon^{\eta}$ centered at $\vec{r}_j$, we also have
\beq\label{chikin}
    \| \nabla \cut \|_2^2 \leq \frac{1}{\vare^{2\eta}} \int_{\cup_{j \in \latt} \bi} \diff \vec{r} \: \left| \nabla |\vec{r}-\vec{r}_j| \right|^2 \leq C N_{\vare} \leq \frac{\constd}{\vare}.
\eeq
The evaluation of the GP functional on $ \trial $ gives
\bmlt
    \label{evaluation}
    \gpf[\trial] \leq C_1 \int_{\mathbb{R}^2} \diff\vec{r} \: \left| \nabla \sqrt{\modu} \right|^2 + C_2 \int_{\mathbb{R}^2} \diff\vec{r} \: \left| \nabla \cut \right|^2 + \int_{\mathbb{R}^2} \diff \vec{r} \: \modu \cut^2 \: \left| \left( \nabla - i \vec{A}_{\varepsilon} \right) \phase \right|^2 + \frac{\tff [ |\trial|^2]}{\varepsilon^2} \\
    \leq \int_{\mathbb{R}^2} \diff \vec{r} \: \modu \cut^2 \: \left| \left( \nabla - i \vec{A}_{\varepsilon} \right) \phase \right|^2 + \frac{\tff [ |\trial|^2]}{\varepsilon^2} + \frac{\constd}{\varepsilon}
\end{multline}
where we have used the bounds \eqref{TFkin} and \eqref{chikin} for the kinetic energies of $ \sqrt{\modu} $ and $ \cut $.
\newline
We can split the first term in \eqref{evaluation} into the contributions from $ \mathcal{B}_{2\rf} $ and its complement respectively. Moreover, exploiting the pointwise estimate for $ r \geq 2\rf $,
\beq
    \label{pointwise gradient}
        \left| \left( \nabla - i \vec{A}_{\vare} \right) \phase \right| \leq \left| \nabla \phase \right| + \frac{\const}{\varepsilon} \leq \sum_{j \in \latt} \frac{1}{|\vec{r} - \vec{r}_j|} + \frac{\const}{\varepsilon}  \leq \frac{\constd}{\varepsilon^{3/2}},
\eeq
and the exponential smallness \eqref{exponential smallness}, one has
\bdi
    \int_{r \geq 2\rf} \diff\vec{r} \: \modu \cut^2 \: \left| \left( \nabla - i \vec{A}_{\varepsilon} \right) \phase \right|^2 \leq \frac{\constd}{\varepsilon^8} \int_{2\rf}^{\infty} \diff r r \: \exp \left\{ - \frac{r - \rf}{\varepsilon} \right\} \leq \frac{\constd }{\varepsilon^6} \exp \left\{ - \frac{\rf}{\varepsilon} \right\}.
\edi
The remaining contribution can be estimated exactly as in \cite{CDY} (see the proof of Theorem 2.1):
\bdi
    \int_{r \leq 2\rf} \diff \vec{r} \: \modu \cut^2 \: \left| \left( \nabla - i \vec{A}_{\varepsilon} \right) \phase \right|^2 \leq C_1 \int_{\lam} \diff \vec{r} \: \left| \left( \nabla - i \vec{A}_{\varepsilon} \right) \phase \right|^2 + C_2 \int_{\cup_{j \in \latt} \bi} \diff \vec{r} \: \cut^2 \left| \left( \nabla - i \vec{A}_{\varepsilon} \right) \phase  \right|^2,
\edi
and an estimate similar to \eqref{pointwise gradient} yields, for $ \vec{r} \in \bi $,
\bdi
    \left| \nabla g_{\varepsilon} \right| \leq \sum_{k \in \latt} \frac{1}{|\vec{r} - \vec{r_k}|} \leq \frac{1}{|\vec{r} - \vec{r_j}|} + \frac{N_{\varepsilon}}{\inf_{j \neq k} |\vec{r}_j - \vec{r}_k|} \leq \frac{1}{|\vec{r} - \vec{r_j}|} + \frac{N_{\varepsilon}}{\spac},
\edi
so that
\begin{multline*}
    \int_{\cup_{j \in \latt} \bi} \diff \vec{r} \: \cut^2 \left| \left( \nabla - i \vec{A}_{\varepsilon} \right) \phase  \right|^2  \leq  2 \int_{\cup_{j \in \latt} \bi} \diff\vec{r} \: \cut^2 \left| \nabla \phase \right|^2 + 2 \int_{\cup_{j \in \latt} \bi} \diff\vec{r} \: \left| \magnp \right|^2 \\
    \leq \frac{C_1\left| \cup_{j \in \latt} \bi \right|}{\vare^{2\eta}} + \frac{C_2N_{\vare}^3}{\vare^{1-2\eta}} + \frac{C_3 N_{\vare}}{\vare^{2-2\eta}} \leq \frac{\constd}{\vare}.
\end{multline*}
The bound \eqref{evaluation} becomes then
\beq
    \label{ela}
    \gpf[\trial] \leq C \int_{\lam} \diff \vec{r} \: \left| \left( \nabla - i \vec{A}_{\varepsilon} \right) \phase \right|^2 + \frac{\tff [ |\trial|^2]}{\varepsilon^2} + \frac{\constd}{\varepsilon} .
\eeq
We now observe  that the upper bound
estimate for the first term in the r.h.s of the above expression can be simply taken over
from Theorem 3.1 in \cite{CDY}: A simple rescaling by $ 2\rf $ immediately yields
\bdi
        \int_{\lam} \diff \vec{r} \: \left| \left( \nabla - i \vec{A_{\varepsilon}} \right) \: g_{\varepsilon} \right|^2 \leq 4\rf^2 \left(\frac{\pi}{2\varepsilon^2}\left(\frac{\omega_0}{2} - \frac{\pi}{\delta^2} \right)^2 + \frac{C_{\omega_0,\delta} |\log\varepsilon|}{\varepsilon}\right),
\edi
and therefore, choosing $ \delta = \sqrt{\frac{2\pi}{\omega_0}} $,
\beq
    \label{phaseeshom}
    \int_{\lam} \diff \vec{r} \: \left| \left( \nabla - i \vec{A_{\varepsilon}} \right) \: g_{\varepsilon} \right|^2 \leq \frac{C_{\omega_0,\delta} |\log\varepsilon|}{\varepsilon}.
\eeq
For the upper bound of the second term in \eqref{ela} we proceed as in \cite{LSY}: Denoting $ W(r) \equiv r^s - \omega_0^2 r^2 / 4 $, one has
\bdi
            \tff[|\trial|^2] - \tfe  \leq \tff[\modu] - \tff[\tfm] + o(\vare) \leq \int_{\mathbb{R}^2} \diff \vec{r}  \: \tfm(r) \left[ j_{\vare} \star W - W \right](r) + o(\vare)
\edi
that is easily estimated using  $|W(|\vec{r} - \vare \vec{r}^{\prime}|) - W(r)|\leq \varepsilon r' C(1+{r}^{s-1})$:
\begin{equation}
    \left[ j_{\vare} \star W - W \right](r) = \frac{1}{2\pi} \int_{\mathbb{R}^2} \diff \vec{r}^{\prime} \: \left[ W(|\vec{r} - \vare \vec{r}^{\prime}|) - W(r) \right] e^{-r^{\prime}}
    \leq \varepsilon C'  (1 + r^{s-1}).
\end{equation}
We thus obtain the estimate $ \tff[|\trial|^2] - \tfe \leq \const \vare $ and
together with \eqref{phaseeshom} this finally yields the upper bound for the GP energy, i.e., $ \vare^2 \gpe \leq \vare^2 \gpf[\trial] \leq \tfe + C_{\omega_0} \vare |\log \vare| $.
\end{proof1}

\begin{proof1}{Corollary \ref{profilehom}}
Defining $ 2 a(r) \equiv \mu^{\rm TF}-r^s+(\omega_0^2/4)r^2$ for all $r\geq 0$ and using the negativity of $a(r)$ outside the support of $ \tfm $, we have
\bdi
    \int_{\mathbb{R}^2} \diff \vec{r} \: (|\gpm|^2 - \tfm)^2 \leq \int_{\mathbb{R}^2} \diff \vec{r} \: \left[ |\gpm|^4 - 2 a(r) |\gpm|^2 + (\tfm)^2 \right] = \tff[|\gpm|^2] - \tfe
\edi
since $ \| \tfm \|_2^2 = \chemtf - \tfe$. The inequality $ \tff[|\gpm|^2] \leq \vare^2 \gpe $ and Theorem \ref{energyhom} thus imply the result.
\end{proof1}

Using Theorem \ref{energyhom} and Corollary \ref{profilehom} we can now show that the density of the minimizer $\gpm$ is
actually exponentially small outside the support of $\tfm$.

\begin{proof1}{Theorem \ref{pom}}
The bound \eqref{denexp} can be derived similarly to Proposition 2.4 in \cite{CDY} or Proposition 2 in \cite{afta}. We present here only the proof of the first statement, since the second one is obtained exactly in the same way.
\newline
The variational equation satisfied by $\gpm$ is \bdi
            - \Delta \gpm - \frac{\omega_0}{\varepsilon} L \gpm + \frac{2}{\varepsilon^2} |\gpm|^2 \gpm +
            \frac{r^s}{\vare^2}\gpm = \chem \gpm
        \edi
where the GP chemical potential $\chem$ is fixed by
$ \|\gpm \|_2 = 1$.  Setting \( U_{\varepsilon}
\equiv | \gpm |^2 \) and using
\beq
    \label{omest}
    \frac{\omega_0}{\vare}\left| {\gpm}^* L \gpm \right| \leq |\nabla \gpm |^2 + \frac{\omega_0^2 r^2 |\gpm|^2}{4\vare^2}
\eeq
we get
\bdi
            -\frac{1}{2} \Delta U_{\varepsilon} \leq \left[ \frac{\omega_0^2 r^2}{4} +
            \varepsilon^2 \chem -r^s - 2 U_{\varepsilon} \right] \frac{U_{\varepsilon}}{\varepsilon^2}.
\edi
The definition of $ \chem $, Theorem \ref{energyhom} and Corollary \ref{profilehom} imply
\begin{multline}
    \vare^2 \chem = \vare^2 \gpf[\gpm] + \int_{\mathbb{R}^2} \diff \vec{r} \: |\gpm|^4 \leq \tfe + C_{\omega_0}\vare |\log \vare| + \int_{\mathbb{R}^2} \diff \vec{r} \: |\gpm|^4 \\
    = \chemtf + C_{\omega_0}\vare |\log \vare| + \int_{\mathbb{R}^2} \diff \vec{r} \: \left[|\gpm|^4 - (\tfm)^2\right]  \leq \chemtf + \const \sqrt{\vare |\log \vare|}.
\end{multline}
On the other hand, since $ a^{\prime}(\rf) < - \const < 0 $ (see Eq.\ (\ref{fprime})), a simple Taylor expansion of $ a(r) $ in a neighborhood of $ \rf $ yields
\bdi
    a(\rf + \vare^{1/3}/2) \leq - \const \vare^{1/3} + \mathcal{O}(\vare^{2/3}) \leq - \const^{\prime} \vare^{1/3}
\edi
for a possibly different constant $ \const^{\prime} > 0 $. Hence
\beq
    \label{unu}
    - \vare^2 \Delta U_{\varepsilon} \leq 2 \left[ 2 a(r) + \const \sqrt{\vare |\log \vare|} \right] U_{\varepsilon} \leq C a(r) U_{\vare} < - \const \varepsilon^{1/3} U_{\vare} < 0
\eeq
for any $ \vec{r} \in \Theta_{\vare} \equiv \{ \vec{r} \in \mathbb{R}^2 \: | \: r > \rf + \vare^{1/3}/2 \} $ and $ \vare $ sufficiently small. Thus $U_{\vare}$ is subharmonic in $\Theta_{\vare}$, so that, for any $ \vec{r} $ and $\varrho$ with $\mathcal{B}_{\varrho}(\vec{r}) \subset \Theta_{\vare} $,
\bdi
    U_{\vare}(\vec{r}) \leq \frac{1}{\pi \varrho^2} \int_{\mathcal{B}_{\varrho}(r)}  \diff \vec{r} \: U_{\vare} \leq \frac{1}{\sqrt{\pi} \varrho} \left[ \int_{\mathcal{B}_{\varrho}(r)} \diff \vec{r} \: U^2_{\vare} \right]^{1/2} \leq \frac{1}{\sqrt{\pi} \varrho} \left[ \int_{r \geq \rf} \diff \vec{r} \: U^2_{\vare} \right]^{1/2} \leq \frac{C_{\omega_0} \sqrt{\vare |\log \vare|}}{\varrho}
\edi
where we have used \eqref{outside support}. If now we take $\vec{r} \in \detfout $ and choose $ \varrho = \vare^{1/3}/2 $, we have
\bdi
    U_{\vare}(\vec{r}) \leq C_{\omega_0}\vare^{1/6} \sqrt{|\log \vare|}
\edi
so that $U_{\vare}(\vec{r}) $ converges pointwise to 0 in $ \detfout $. Moreover from \eqref{unu} it follows that $U_{\vare}$ is a subsolution in $ \detfout $ of
\beq
    \label{her}
        \left\{
            \begin{array}{l}
                    - \Delta w + \const \vare^{-5/3} w = 0 \\
                    \mbox{} \\
                    w(\partial \detfout) = C_{\omega_0}\vare^{1/6} \sqrt{|\log \vare|},
            \end{array}
            \right.
\eeq
whereas the r.h.s. of \eqref{denexp} is a supersolution of the same problem for $ \vare $ sufficiently small. The result is then a consequence of the comparison principle.
\end{proof1}

\subsection{The regime $\omega(\vare) \gg 1/\vare$}
\label{proofsultra}
\begin{proof1}{Theorem \ref{energyahom}}
    In order to capture the leading order term in the GP energy asymptotics it is convenient to rescale the GP functional in the following way: Setting $ \eneg[\tilde{\Psi}] \equiv \vare^2 \rmax^{-s} \gpf[\Psi] $, with  $ \tilde{\Psi}(\vec{x}) \equiv \rmax \Psi(\vec{r}) $ and $ \vec{x} \equiv \rmax^{-1} \vec{r} $, we have (remember that $ \rmax $ depends on $ \vare $ through $\omega_0(\varepsilon)=\omega_1/\varepsilon^\alpha$)
    	\beq
    	\label{rescaledultra}
        	\eneg[\tilde{\Psi}] = \vare^2 \rmax^{-(s+2)} \int_{\mathbb{R}^2} \diff \vec{x} \: \left| \left( \nabla - i \vec{\mathcal{A}}_{\vare} \right) \tilde{\Psi} \right|^2 + \enef[|\tilde{\Psi}|^2],
    	\eeq
	where $\enef $ is the TF functional \eqref{TFfastrot} with $ \omega_0 $ replaced with $ \omega_1/\vare^{\alpha} $ and
	\bdi
		\vec{\mathcal{A}}_{\vare} \equiv \frac{\omega(\vare)\rmax^2}{2} \: \vec{e}_z \times \vec{x}.
	\edi
    The proof is thus similar to that of Proposition \ref{TFenergyasympt}: By neglecting the (positive) first term in \eqref{rescaledultra} we get the lower bound
    \beq
        \eneg[\tilde{\Psi}] \geq 1 - \frac{s}{2}.
    \eeq
    In order to obtain a corresponding upper bound we test the functional on a trial function $ \trialultra $ similar to the one used in the proof of Proposition \ref{TFenergyasympt}, i.e.,
    \beq
        \trialultra(\vec{x}) \equiv \sqrt{\tilde{\rho}_{\xi}(x)} \exp \left\{ i \left[ \frac{\omega(\vare) \rmax^2}{2} \right] \vartheta \right\}
    \eeq
    where $ [ \:\cdot\: ] $ stands for the integer part and the density $ \tilde{\rho}_{\xi}(x) $ is defined in \eqref{trialdensity} (we additionally require that $ \| \nabla \sqrt{j} \|_2 < \infty $).
    \newline
    The estimate of the second term in \eqref{rescaledultra} is thus already done in \eqref{estimateTFultra}. It remains to bound the kinetic energy of $ \trialultra $, i.e., 
    \bdi
        \int_{\mathbb{R}^2} \diff \vec{x} \: \left| \left( \nabla - i \vec{\mathcal{A}}_{\vare} \right) \trialultra \right|^2 = \int_{\mathbb{R}^2} \diff \vec{x} \left| \nabla \sqrt{\tilde{\rho}_{\xi}} \right|^2 + \int_{\mathbb{R}^2} \diff \vec{x} \: \left\{ \frac{1}{x} \left[ \frac{\omega(\vare) \rmax^2}{2} \right] - \frac{\omega(\vare) \rmax^2  x}{2} \right\}^2 \tilde{\rho}_{\xi}(x).
    \edi
    Smoothness of $ \tilde{\rho}_{\xi} $ (and the assumption $ \| \nabla \sqrt{j} \|_2 < \infty $) yields the estimate
    \bdi
         \int_{\mathbb{R}^2} \diff \vec{x} \left| \nabla \sqrt{\tilde{\rho}_{\xi}} \right|^2 \leq C \xi^{-2}
    \edi
    while, for $ \xi $ sufficiently small,
    \begin{multline*}
        \int_{\mathbb{R}^2} \diff \vec{x} \: \left\{ \frac{1}{x} \left[ \frac{\omega(\vare) \rmax^2}{2} \right] - \frac{\omega(\vare) \rmax^2  x}{2} \right\}^2 \tilde{\rho}_{\xi}(x) \leq \omega^2(\vare) \rmax^4 \int_{\mathbb{R}^2} \diff \vec{x} \: \frac{(1-x^2)^2}{x^2} \tilde{\rho}_{\xi}(x) + \int_{\mathbb{R}^2} \diff \vec{x} \: \frac{\tilde{\rho}_{\xi}(x)}{x^2} \leq   \\
        C_1 \omega^2(\vare) \rmax^4 \xi^{-1} \int_{1-\xi/2}^{1+\xi/2} \diff z \: \frac{(1-z)^2}{z} j(\xi^{-1}(1-z)) + C_2 \leq C \omega^2(\vare) \rmax^4 \xi^{2}.
    \end{multline*}
    Altogether we get the bound
    \beq
        \eneg[\trialultra] \leq 1 - \frac{s}{2} + \mathcal{O}(\xi^2) + \mathcal{O}\left(\vare^{\frac{2\alpha(s+2)}{s-2}} \xi^{-1} \right) + \mathcal{O} \left(\vare^2 \vare^{\frac{2\alpha(s+2)}{s-2}} \xi^{-2}\right).
    \eeq
    Optimizing with respect to the first two error terms we obtain the same error term as in \eqref{tfentildeasympt} and the last term gives only a higher order correction, as long as $ \alpha \leq \frac{3(s-2)}{s+2} $. On the other hand, for larger $ \alpha $, we consider the first and last terms in the above estimate and choose (in this case the second term can be neglected)
    \bdi
        \xi = \sqrt{\vare} \: \vare^{\frac{\alpha(s+2)}{2(s-2)}}
    \edi
    which yields an overall remainder of order $ \vare \: \vare^{\frac{\alpha(s+2)}{s-2}} $.
    \end{proof1}
    \begin{proof1}{Theorem \ref{expahom}}
The first part of Theorem \ref{expahom} can be proved exactly as
in the proof of Theorem \ref{pom} and we omit the details. The
weak convergence to a Dirac delta function supported at $ x =1 $ is
a simple consequence of the pointwise estimate \eqref{pointestultra}
together with the $ L^1-$normalization of the density $ |
\tgpm(\vec{x}) |^2 $ (see, e.g., the discussion in Section
\ref{support infinite}).
\end{proof1}

\section{Asymptotically homogeneous potentials}

We recall from the Introduction  that a potential 
 $V(r)$ is called asymptotically homogeneous if there are constants $\kappa,c>0$ such that  the estimate
 \beq
	\label{homogeneous2}
	\left| \lambda^{-s} V(\lambda r) - r^s \right| \leq c \lambda^{-\kappa} \left( 1+ r^s \right)
\eeq holds for all  $\lambda\geq 1$ and  all  $r\in\mathbb R^+$. We discuss here briefly how the results for the trapping potential $r^s$ can be extended to such  potentials $V(r)$ with suitable modifications of the error terms.

The rescaling that produced  \eqref{scaling}  leads in the general case to 
\beq 
	\label{gpfu}
 	\gpfu[\Psi] = \int_{\mathbb{R}^2} \diff \vec{r} \: \left\{ |\nabla \Psi|^2 - \omega(\vare) \Psi^* L \Psi+ \frac{|\Psi|^2}{\vare^2} \left[ \vare^{\frac{2s}{s-2}} V \left(\vare^{-\frac{2}{s-2}}r \right) + |\Psi|^2 \right]  \right\},
\eeq
i.e., the functional contains the rescaled external potential $ \lambda^{-s} V(\lambda r) $ with $ \lambda = \vare^{-\frac{2}{s-2}} $.

The estimates mentioned in Subsection 3.1  for $\omega\ll 1/\varepsilon$ generalize to asymptotically homogeneous potentials in exactly the same way as in \cite{LSY}. For the case $\omega\sim 1/\varepsilon$ we have

\begin{pro}[GP energy and density asymptotics for $\omega\sim 1/\varepsilon$]
\label{asympthompotentials}
\mbox{}	\\
Let the external potential $ V(r) \geq 0 $ be asymptotically homogeneous of degree $ s > 2 $ in the sense of \eqref{homogeneous} and let $ \gpeu $ and $ \gpmu $ denote the ground state energy and wave function of the functional \eqref{gpfu}. 
\newline
Then for any $ \omega_0, \kappa > 0 $ fixed and $ \vare $ tending to $0$,
\beq
	\label{asympthompotene}
	\varepsilon^2 \: \gpeu = \tfe + \mathcal{O}(\varepsilon | \log\varepsilon |) + \mathcal{O}\left( \vare^{\frac{2\kappa}{s-2}} \right).
\eeq
Furthermore the density $ | \gpmu |^2 $ converges to $ \tfm $ strongly in $ L^1(\err) $.
\end{pro}

\begin{proof}
The proof requires only a minor modification of the proof of Theorem \ref{energyhom}:
Using \eqref{homogeneous}, we can estimate
\beq
	\vare^2 \left| \gpfu\left[ \Psi \right] - \gpf\left[\Psi \right] \right| \leq c \: \vare^{\frac{2\kappa}{s-2}} \int_{\err} \diff \vec{r} \: (1+r^s) \left| \Psi \right|^2,
\eeq
so that the appropriate upper and lower bounds to $ \gpeu $ can be easily obtained: By testing the functional on $ \gpm $ we immediately get the upper bound $ \gpeu \leq \gpe + o(1) $, whereas taking $ \Psi = \gpmu $ in the above inequality, one has the lower bound
\beq
	\vare^2 \gpeu \geq \vare^2 \gpe - c \: \vare^{\frac{2\kappa}{s-2}} \int_{\err} \diff \vec{r} \: r^s  \left| \gpmu \right|^2,
\eeq  
which yields the expected result, provided one can show that there exists a finite constant $ \const $, such that
\beq
	\label{uboundpotential}
	\int_{\err} \diff \vec{r} \: r^s  \left| \gpmu \right|^2 \leq \const.
\eeq
On the other hand, evaluating $ \gpfu $ on a smooth radial function, we see that $ \vare^2 \gpeu \leq \const^{\prime} $, for some finite constant $ \const^{\prime} $, so that
\bdi
	\int_{\err} \diff \vec{r} \: \left[ \vare^{\frac{2s}{s-2}} V \left(\vare^{-\frac{2}{s-2}} \vec{r} \right) - \frac{\omega_0^2 r^2}{4} \right] \left| \gpmu \right|^2 \leq \const^{\prime}
\edi
but, using the trivial bound ($ r_{0} \equiv (\omega_0^2/2)^{\frac{1}{s-2}} $), 
\begin{multline*}
	\int_{\err} \diff \vec{r} \: r^2 \left| \gpmu \right|^2 \leq \left( \frac{\omega_0^2}{2} \right)^{\frac{2}{s-2}} \int_{r \leq r_{0}} \diff \vec{r} \: \left| \gpmu \right|^2 + \frac{1}{2} \int_{r \geq r_{0}} \diff \vec{r} \: r^s \left| \gpmu \right|^2 \leq \\
	\left( \frac{\omega_0^2}{2} \right)^{\frac{2}{s-2}} + \frac{1}{2} \int_{\err} \diff \vec{r} \: r^s \left| \gpmu \right|^2,
\end{multline*}
together with \eqref{homogeneous}, we get \eqref{uboundpotential}, i.e.,
\bdi
	\left( \frac{1}{2} - o(1) \right) \int_{\err} \diff \vec{r} \: r^s  \left| \gpmu \right|^2 \leq \const^{\prime} + \frac{\omega_0^2}{4} \left( \frac{\omega_0^2}{2} \right)^{\frac{2}{s-2}} + o(1).
\edi
The energy asymptotics follows then from Theorem \ref{energyhom}.
\newline
In order to prove the ground state density convergence, it is sufficient to note that \eqref{asympthompotene} implies that $ |\gpmu|^2 $ is a minimizing sequence for the TF functional $ \tff $. The statement can be thus obtained by a simple compactness argument together with identity of norms, $ \left\| \gpmu \right\|_2^2 = \left\| \tfm \right\|_1 = 1 $ (see, e.g., Theorem II.2 in \cite{LSY2}).
\end{proof}

For $\omega(\varepsilon)=\omega_1/\varepsilon^{1+\alpha}$ we have the following generalization of Theorems \ref{energyahom} and \ref{expahom}:
\begin{pro}[GP energy and density asymptotics for $\omega\gg 1/\varepsilon$]
\label{ultraasymhompot}
\mbox{}	\\
Let the external potential $ V $ satisfy the same conditions as in Proposition \ref{asympthompotentials}. 
\newline
Then for any fixed $ \omega_1, \alpha, \kappa > 0 $ and $ \vare $ tending to zero,
\beq
	 \varepsilon^{2} \: \vare^{\frac{2\alpha s}{s-2}} \: \gpeu = \left( \frac{\omega_1^2}{2s} \right)^{\frac{s}{s-2}} \left( 1 - \frac{s}{2} \right)  + \mathcal{O} \left( \varepsilon^{\frac{4\alpha(s+2)}{3(s-2)}} \right) + \mathcal{O} \left( \vare \: \varepsilon^{\frac{\alpha(s+2)}{s-2}} \right) + \mathcal{O} \left( \vare^{\frac{2\kappa}{s-2}} \vare^{2\alpha \kappa} \right).
\eeq
Furthermore the rescaled density $ \big| \tilde{\Psi}^{\mathrm{GP}}_{\vare,V} (\vec{x}) \big|^2 \equiv \rmax^2 \big| \gpmu(\rmax \vec{x}) \big|^2 $ converges in the sense of distributions to a Dirac delta function concentrated at $ x =1$. 
\end{pro}

\begin{proof}
It is sufficient to rescale the functional \eqref{gpfu} as in \eqref{rescaledultra}, i.e., setting $ \eneg_V [\tilde{\Psi}] \equiv \vare^2 \rmax^{-s} \gpfu[\Psi] $,
\begin{multline*}
	\eneg_V [\tilde{\Psi}] = \vare^2 \rmax^{-(s+2)} \int_{\mathbb{R}^2} \diff \vec{x} \: \left| \left( \nabla - i \vec{\mathcal{A}}_{\vare} \right) \tilde{\Psi} \right|^2 + \\
	\int_{\err} \diff \vec{x} \: \left\{ \left[ \rmax^{-s} \vare^{\frac{2s}{s-2}} V \left( \vare^{-\frac{2}{s-2}} \rmax x \right) - \frac{sx^2}{2} \right] |\tilde{\Psi}|^2 + \rmax^{-s-2} | \tilde{\Psi} |^4 \right\}
\end{multline*}
and proceed as in the proof of Proposition \ref{asympthompotentials} to get the estimate
\bdi
	\vare^2 \rmax^{-s} \left| \tilde{E}^{\mathrm{GP}}_{\vare,V} -  \tilde{E}^{\mathrm{GP}}_{\vare} \right| \leq \constultra \vare^{\frac{2\kappa}{s-2}} \rmax^{-\kappa}. 
\edi
Theorem \ref{energyahom} now yields the result.
\newline
The rescaled density $ \big| \tilde{\Psi}^{\mathrm{GP}}_{\vare,V} (\vec{x}) \big|^2 \equiv \rmax^2 \big| \gpmu(\rmax \vec{x}) \big|^2 $ converges in the sense of distributions to a Dirac delta function concentrated at $ x =1$ by the same arguments as before (cf.\  the proof of Theorem \ref{expahom} in Section \ref{proofsultra}). 
\end{proof}

\section{Conclusion} We have analyzed in some detail the TF limit of the GP energy and density of a rapidly rotating Bose Einstein condensate in a two-dimensional trapping potential of the form $r^s$, $s>2$.
After discussing the scaling of the variables (that is necessary because of spreading due to the interaction and centrifugal forces)  we have estimated the energy with error terms whose order in the small parameter can be expected to be optimal and  proved the concentration of the GP density on the  support of the TF density apart from exponentially small terms. The extension to asymptotically homogeneous potentials and the corresponding change of the error terms has also been  discussed.

\vspace{1cm}
\noindent{\bf Acknowledgments:}  This work was supported by the Austrian Science Fund (FWF) grant P17176-N02 and the EU Post Doctoral Training Network HPRN-CT-2002-00277 ``Analysis and Quantum''. JY gratefully acknowledges hospitality at the Institute Henri Porincar\'e in Paris during the trimester \lq Gaz quantiques' 2007.

\end{document}